%% file: MTO.tex
\documentclass[12pt,authoryear,times,5p,twocolumn,url=false]{elsarticle}
\usepackage{amssymb}
\usepackage{amsmath}
\usepackage{graphicx}
\usepackage{epstopdf}
\usepackage{booktabs}
\usepackage{topcapt}
\usepackage{setspace} 
\usepackage{examplep}

\usepackage{pst-all}
\usepackage{setspace}

\usepackage{url}
\usepackage{txfonts}
\usepackage{algorithm}
\usepackage{algpseudocode}

\newtheorem{example}{Example}[section]

\providecommand{\bred}   {\ensuremath{\textcolor[rgb]{1,0,0}  {\bullet}}}
\providecommand{\bgreen} {\ensuremath{\textcolor[rgb]{0,.5,0} {\bullet}}}
\providecommand{\borange}{\ensuremath{\textcolor[rgb]{1.,.4,0}{\bullet}}}

\journal{arXiv}

\begin{document}
\title{Splitting hybrid Make-To-Order and Make-To-Stock demand profiles}

\author{Wolfgang Garn,James Aitken}
\address{The Surrey Business School, University of Surrey, Guildford, Surrey, GU2 7XH, United Kingdom}
\ead{\{w.garn,james.aitken\}@surrey.ac.uk}
\tnotetext[phone_and_fax_number]{\\ Tel.: +44(0)1483 68 2005; fax: +44(0)1483 68 9511.}

\begin{abstract}
 
 
In this paper a demand time series is analysed to support Make-To-Stock (MTS) and Make-To-Order (MTO) production decisions. Using a purely MTS production strategy based on the given demand can lead to unnecessarily high inventory levels thus it is necessary to identify likely MTO episodes.

This research proposes a novel outlier detection algorithm based on special density measures. We divide the time series' histogram into three clusters. One with frequent-low volume covers MTS items whilst a second accounts for high volumes which is dedicated to MTO items. The third cluster resides between the previous two with its elements being assigned to either the MTO or MTS class. The algorithm can be applied to a variety of time series such as stationary and non-stationary ones. 

We use empirical data from manufacturing to study the extent of inventory savings. The percentage of MTO items is reflected in the inventory savings which were shown to be an average of 18.1\%.

\end{abstract}
\begin{keyword}
	demand analysis; time series; outlier detection; production strategy; Make-To-Order(MTO); Make-To-Stock(MTS); 
\end{keyword}

\maketitle

\section{Introduction}

Research into production systems has generally characterized and modeled them as either make-to-order (MTO) or make-to-stock (MTS). We propose the following definitions. A \textit{make-to-order product }is a product where the required quantity is manufactured after receiving a sales order. MTO products are commonly identified by low average demand and a high coefficient of variation \citep{Soman2007}. A \textit{make-to-stock product} is a product where items are manufactured on anticipated demand. A concrete sales order may not exist. Usually the required quantity is derived from forecasts.

Presenting the operating choice of production systems as a choice between MTO or MTS simplifies the discussion but does not reflect current production dynamics. Fewer and fewer firms can be classified as purely MTS or MTO in practice \citep{Christopher2010, Soman2006, Aitken2003}. Investigating and modeling the combined MTS-MTO challenge has had limited research with only a few papers exploring some aspects of the problem \citep{Soman2006,Rajagopalan2002}. 
Where capacity exists for manufacturers it can be expedient to separate products between MTO and MTS in terms of planning and control. The literature highlights the benefits of a focusing/isolating production of MTO and MTS in terms of changeovers, bottleneck reduction, process variation and costs \citep{Schmenner2004}. However, separating products between the two approaches is not possible for many firms. Sharing and scheduling finite capacity between MTS and MTO products is a major challenge for many organisations \citep{Kerkkaenen2007}. This paper investigates how organisations manage the MTS-MTO dynamic in the context of a food manufacturing business. The food sector predominately services supermarkets and is under growing pressure to increase the number of stock keeping units (SKUs), reduce lead-times and manage the delivery of an increasing heterogeneous service demand across customers.  The growing complexity of operating in the food sector has increased the incidences of shared capacity between MTO and MTS. The sector offers the opportunity of investigating the combined MTS-MTO scheduling challenge in the context of finite capacity faced with high demand variability products classified as either MTO or MTS. 

Within the food sector companies arrange bulk sales, promotions or marketing events that impact on the demand pattern of products. These events are not always visible to the production planners leading to unplanned surges in demand creating unstable production schedules and diminishing service levels. Unplanned surges can lead to product being purely produced via a MTS process when they actually require a MTO production strategy. Ignoring this often results in the depletion of MTS produced inventory for regular sales and subsequently unplanned changes in production \citep{Schmenner2004}.  We introduce a new approach, which identifies MTO quantities within an otherwise MTS classified product. The objective of our paper is to present a method that can support the identification of changes in status between MTS and MTO. Identification of changes can motivate a MTO strategy for products which are treated as MTS, a hybrid approach, providing the advantage of a production schedule which reduces inventory and improves customer service. The case study demonstrates that applying our new method led to reduced inventory levels of 18.1\% on average for previously pure MTS classified products. The proposed new method automatically identifies MTO items as ``outliers" within a time series, allowing the absence of order records. The novelty of the method is the usage of clustering demand frequency extracted from time series.

Our paper is organized as follows; after introducing the case study firm we review the literature on MTS-MTO and MTO/MTS in the section three and then turn in section four to the new proposed method for the identification of change in product status between MTS, MTO and MTO/MTS. Section five presents the results from the modeling and implementation of the new approach in the case study firm followed by discussions on the significance of identifying and managing the status change between MTO and MTS. Section six presents the conclusions and suggestions for future work in this area of growing importance for operation managers.

\section{Company and Process Overview\label{sec:company}}
The case study firm produces over 250 different products derived from five raw materials. The raw materials are cleaned and sorted leading to the production of eight intermediary semi-finished products which are stored in material handling silos awaiting release to the appropriate production lines for packing. Storage of the semi finished products was limited to 48 hours due to bacterial and decay issues. The variety of different packaging options generated in excess of 250 finished products for sale to the retail and catering market segments. The production process for the food processing business is shown in figure 1, 
\begin{figure}[]
	\centering
		\includegraphics[width=1.00\columnwidth]{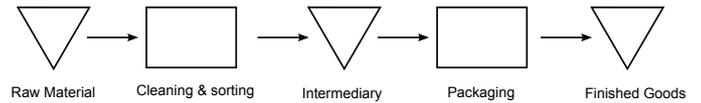}
	\caption{Production process.}
	\label{fig:production-process}
\end{figure}
The manufacturing process had been faced with the challenges of growing sales, increased supply chain complexity and falling service levels of a broadening heterogeneous customer base.  The result of the increasingly variable demand on the manufacturing operation was the conversion of the production and planning control schedule from a plan with a three week time horizon to a reactive customer order list. The reactive planning and control modus operandi had forced manufacturing into an increasing number of unplanned changeovers and diminishing productivity, effectively reducing the available capacity in the packaging operation. The firm had moved from a position of flexible spare capacity to a bottleneck over a three year period due to increased complexity and reduction in schedule stability. 

The case study company reflects the  food manufacturing sector which has come under increasing pressure from supermarkets and distributors to diminish the homogenisation of their offerings, reduce lead-times, increase promotions, shorten product life-cycles and provide more frequent deliveries \citep{Fisher1999,Taylor2006, Squire2009}. The continually changing environment that the food manufacturers operate within has made the management of processes through the application of forecasting methods difficult. The inaccuracies of forecasting in the food sector have been found to have a significant impact on the performance and efficiency of the operation \citep{Taylor2006,Aghazadeh2004}. Manufacturing a large number of products through a pure MTS approach and putting them into inventory is not viable because of unpredictable demand and shelf-life \citep{Soman2007}. Due to increasing volatility the effectiveness and returns from increased investment on forecasting could be viewed as futile \citep{Christopher2010} rendering the pure MTS approach ineffective. In order to improve the performance of the operation the case study firm decided to alter its approach to planning and control of MTO and MTS categorised products 

\section{Literature Review}

\subsection{MTS and MTO}
Make-To-Stock (MTS) systems are described in terms of low variety and high volume that are driven through forecasts \citep{Rafiei2011,Soman2004}. Operationally the issues that are in focus for this system are lot size determination, accuracy of forecasting and inventory control \citep{Soman2006}.  A MTS approach is used for manufacturing items which are standardized with high volume, regular demand and are delivered to customers from stock \citep{Birou2011,Chang2010, Kerkkaenen2007}.  
The preferred option for firms is to move from a MTS to a MTO system of production linked to a change in philosophy from a push to pull approach \citep{Birou2011, Jodlbauer2008}. MTO products are outlined  in terms of customised items with low average volumes and irregular demand \citep{Birou2011, Kerkkaenen2007}.  The uncertainty linked to MTO products places planning and control process in a pivotal and time-sensitive critical role within firms \citep{Corti2006}. Reducing lead-times and improving the reliability of due dates, in the context of demand uncertainty,  has been a focus for several researchers in developing models that can support firms in their decision making to optimize production schedules and performance \citep{Zaerpour2009, Jodlbauer2008, Sawik2006}.  

An alternative approach to managing the pressure related to reducing lead-times and growing uncertainty in demand is through the development of a hybrid MTS/MTO approach. The approach attempts to harness the strengths of the pure MTO and MTS systems \citep{Federgruen1999}. Through operating a MTS system in the downstream sections of the operations parts can be produced that are subsequently assembled in the MTO section when orders are received \citep{Rafiei2011}. Heuristic models have been developed to control and schedule production activities to support the portioning problem and successfully operate the hybrid approach \citep{Chang2003, Rajagopalan2002}. 
The splitting of the operation reflects the concept of ``postponement” which has successfully been applied in many sectors where it is possible to store semi-finished products or modular parts awaiting orders before assembling the final product without any detritus effect on quality \citep{Zaerpour2009, Kerkkaenen2007}. 
The option to build semi-finished product and store the materials until an order is received has limited application in the food processing sector.  Strict food storage guidelines, bacterial control challenges and short shelf-life render the opportunity for utilizing a hybrid MTS/MTO approach, through semi-finished products, difficult if not illegal \citep{Johnston2003}.  However the mixing of the MTO and MTS approaches has merit outside of the ``postponement” concept. Firms in the food sector dynamically move from a MTS to MTO status to manage short term changes in demand operating a hybrid MTS/MTO approach not based on ``postponement".  

MTS/MTO processes enjoy popularity in industry and find themselves implemented in software systems.
Enterprise resource planning systems such as SAP, Microsoft Dynamics and Oracle's JD Edwards accommodate  strategies for MTS and MTO products. For instance SAP has a detailed default process flow for handling MTO products. The user is forced to distinguish a product to be either of MTO or MTS type. That means products of hybrid character lead to two product records, which evolve into two demand and production time series. In this paper we develop an alternative approach that reverses the process by having one demand time series for a hybrid product and splitting it up into MTO and MTS items for production planning and control. Some researchers have begun tackling the challenge of modeling MTS/MTO systems and to analytically decide whether a product should be classified as MTS or MTO through the lens of demand analysis. The methodology section builds on this earlier approach through the application of outlier detection techniques leading to the introduction of the new method developed to dynamically identify MTO and MTS product categories.

\subsection{Outlier detection techniques\label{sec:outlier techniques}}
Demand analysis has been utilised to categorise products into MTS and MTO families.  \cite{D'Alessandro2000} suggested that an analysis through the prism of demand variability and average weekly demand products can be categorised as MTO or MTS. A fair amount of literature in this field utilises the coefficient of variation (CoV):
\begin{equation} c_v = \frac{\sigma}{\mu},\end{equation}
which is the ratio of standard deviation $\sigma$ and mean $\mu$.
 Products with low demand variability and ``high volume" were classed as MTS, remaining products were viewed as MTO. A typical characterisation of a MTO product is its low average demand and high variation.Some researchers have stated that a coefficient of variation on or below 0.5 may be used to classify a product as MTS. The authors have not found any definitive method in the literature, which computes the likeliness of a product falling into the MTS or MTO category.  Determining where the cut-off point is for demand variability and the portioning of products into MTO or MTS based on the CoV factor can be subjective and difficult to determine \citep{Soman2007}.  Recent research has challenged the validity of applying of CoV as a lens to allocate MTS or MTO status due to the frequent changes in CoV for products as market and their demands change quickly altering a product categorization \citep{Godsell2012}. The low average demand could be deceiving because large, unforecasted, orders can shift the average to any level.  The ratio of the sample standard deviation to the sample mean overcomes this problem. However, trends can affect this ratio. Removing the linear trend leads to improvements. 

In practice operations need to derive a production strategy that accommodates MTS and MTO from a single demand profile. We will assume that outliers represent MTO items. Hence, we have to considered outliers in time series. \cite{Lewis1975} reviews demand analysis in his extensive work where he discusses demand impulses in the context of adaptive forecasting He emphasises that ``\textit{unusual demands still have to be dealt with}" indicating the relevance of outlier detection.
\cite{Silver1985} also acknowledge that ``\textit{temporary change in demand pattern}" such as promotions should be filtered out before forecasting.Removing identified outliers will reduce the variation of the MTS classified product thus reducing subsequent inventory.

\cite{Atkinson1997} suggested ways of checking fitted models to possible shocks and introduced a new type of intervention analysis, which uses an extra parameter for the outlier. This is equivalent to deleting the outlier and observing the residual sum of squares in a regression model. A considerable amount of literature about outlier detections refers to \cite{Fox1972}, which introduces likelihood ratios. Fox looks at stationary time series and proposes two types of errors. The first type deals with recording error or ``isolated independent cross execution" that are independent of other observations. The second type of ``outliers" affects consecutive observations - known as additives or innovations. Our work focuses on the first type, which is commonly addressed as shock in the time series literature. Time series analysis usually relies on predictive causality, the opposite of what we aim to do. 

 \cite{Hawkins1980} discusses the identification of outliers in great detail and defines them as: ``\textit{An outlier is an observation which deviates so much from the other observations as to arouse suspicions that it was generated by a different mechanism}". It is convenient to introduce a \textit{basic model}, which is univariate, assumes one ``normal" generating mechanism and that outliers are rare observations. \cite{Barnett1994} similarly define an \textit{outlier} in a set of data to be ``\textit{an observation (or subset of observations) which appears to be inconsistent with the remainder of that set of data}".
On the other hand \textit{contaminants} are observations which are not ``genuine members" of the main distribution. Extremes could be outliers or contaminants. Assume that the actual distribution $F$ is known for a sorted sample $x_1,x_2,\dots,x_n$. $x_1$ and $x_n$ are the sample extremes, which might be outliers. However, outliers are extreme values. Contaminants interfere with the original distribution $F$ and may stem from another distribution $G$. This could result in extreme values wrongly identified as outliers. An intuitive model could assume that a demand time series is made up of a MTS distribution $F$ and a MTO distribution $G$. 

 A variety of methods have been devised to detect outliers. Typically these techniques fall into two categories: model based and proximity based approaches \citep{Kriegel2010}. Model based approaches can use statistical tests, depth considerations and deviation analysis. Proximity based techniques usually use distance measures or density aspects. 
\cite{Chen2010} compare outlier detection techniques in regards of of performance. They analysed statistics-based, distance-based  and density-based approaches.
Model based approaches have been extensively investigated by \cite{Hawkins1980}. Most of these models assume normal distributed data; however, all samples in the case study reject this assumption. We will focus our review on proximity based techniques.  
Distance based approaches have the basic assumption that data has a dense neighbourhood. This allows identification of outliers due to their distance from the neighbourhood. \cite{Knorr1998,Knorr1999} have discussed such algorithms based on iodizes, nested-loops and grids. The $k$-nearest neighbour method was adapted for outlier detection by \cite{Ramaswamy2000}, \cite{Angiulli2002} and many more.
Proximity techniques based on density usually assume that the density around an outlier is different to its neighbours. \cite{Breunig1999,Breunig2000} use the concept of the Local Outlier Factor that overcomes the issue of clusters having different densities. Related to this concept are the works from \cite{Tang2002} and \cite{Jin2001}. A further improvement was achieved by \cite{Jin2006} by using symmetric neighbourhood relationships. 
\cite{Cao2010} developed a density-based algorithm to identify outliers. They introduced a density-similarity-neighbourhood factor which suggests the likeliness of the data to be an outlier.
Another interesting approach which uses a local outlier correlation integral on an $\epsilon$ neighbourhood is \cite{Papadimitriou2003} work.  \cite{Kriegel2010} mentioned that cluster algorithms can be used to identify a noise set/outliers. Multiple outliers similar to each other would form a cluster rather than being identified as noise. 
\cite{Alan2011} use thresholds to form multiple clusters using proximity based outlier detection mechanisms. 
These techniques are insightful but were not designed for demand time series in particular.

\subsection{MTO cluster detection technique\label{sec:CDT}}
In this work we propose a new technique which is based on proximity and density. The novelty of our approach is a distance measure operating on the probability mass distribution that identifies MTS and MTO categories via clustering. Before we embark on the details of this technique we give a first demand taxonomy of hybrid MTS-MTO products:
\begin{enumerate}
	\item MTO outliers 
	\begin{enumerate}
		\item Demand is high and infrequent 
		\item Demand is low and infrequent
	\end{enumerate}
	\item MTO contaminants
	\begin{enumerate}
		\item MTO and MTS demand distributions differ
		\item MTO and MTS demand are similar
	\end{enumerate}
	
\end{enumerate}
Type 1 (a) assumes that MTO items are characterised by high demand and low frequency. This is the case we will develop our techniques for.
Type 1 (b) is not important to us because we assume that MTS production will not be affected significantly as low demand can be served from inventory.
Type 2 (a) allows a separation of the distributions when determination of the parameters is a possibility.
Type 2 (b) constitutes a challenge and cannot be analysed without further information and will be discussed in future research.
 
\section{Methodology}

In this section we will introduce new methods, which help to classify what parts of the demand should be used as MTS or MTO. Some of the theoretical underpinnings were introduced in the previous section, which motivated a discussion of outlier detection techniques. Furthermore we derived a taxonomy of hybrid MTS-MTO systems. 

Here, we will propose our new MTS and MTO identification method. This requires us to introduce a novel series distance measure, which operates on histograms and time series. We embed this distance measure into a cluster technique and derive three categories. A new immediate neighbour detection technique reduces these to the two final MTS and MTO categories.

The methodology introduced in this research focuses on type 1 demand profile. Our goal is to separate the convoluted MTS-MTO demand. The first step is the definition of an appropriate distance metric. A new city block metric that operates on a histogram with frequencies $f=(f_1,\dots,f_n)$ is proposed. 
\begin{equation}
	\breve{d}_{st}=|f_s-f_t|+\alpha |c_s-c_t|,
\end{equation}
The distance $\breve{d}_{st}$ between $s$ and $t$ uses equidistant classes $c=(c_1,\dots,c_n)$ with $\breve{d}=c_{i+1}-c_i>0, i<n$, where $c_i$ represents the centre of class $i$. Note that a scaling factor $\alpha$ was introduced.  
The factor is supported by the following considerations.
Assume each class contains only one observation then the unit distance between classes seems appropriate. If each class contains ten observations a unit distance would be inappropriate, because the class distance is not in proportion to the frequency scale. This motivates the \textit{class scaling factor} $\alpha$: 
\begin{equation}\alpha = \frac{\sum_{k=1}^n f_k}{(n-1)\breve{d}}. \label{eq:alpha}\end{equation}
\begin{example}[scaled city block distance]
\begin{figure*}
 \centering
	\begin{tabular}{cc}
		\includegraphics[width=8cm]{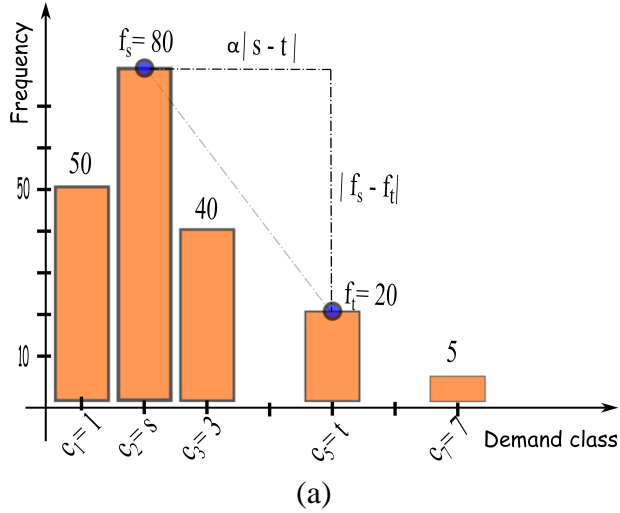} &
		\includegraphics[width=8cm]{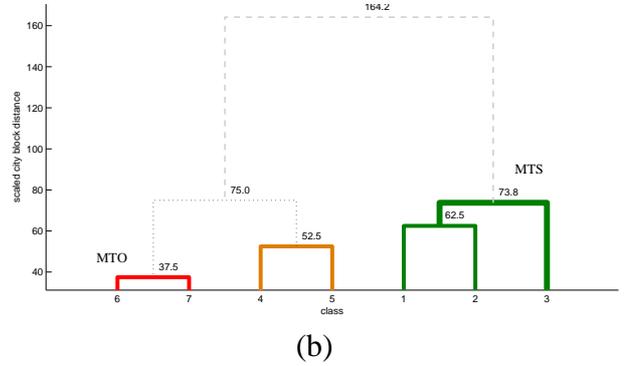} \\
		(a) & (b) 
	\end{tabular}
	\caption{Example (a) demand histogram showing scaled city block metric. (b) dendogram.}
	\label{fig:Example}
\end{figure*}
Let figure \ref{fig:Example} (a) be the histogram derived from a demand time series. Class two shows demand for two items occurs 80 times/days. A total of 195 days were observed and placed into seven demand classes. The scaling factor is $\alpha = 195/(7-1) = 32.5$. Hence the distance between $s=(2,80)$ and $t=(5,20)$ is $\breve{d}_{st} = |80-20|+32.5|2-5|=60+97.5=157.5$. Note that the demand classes can be transformed to represent other quantities (e.g. $c_1$ could represent 1200 items, $c_2=1300$ with $\breve{d}=100$)).
\end{example}
The problem with the above distance measure is that for instance the distance between two classes separated by another one could be shorter than the one next to each other. As a result clusters could consist out of disconnected classes. We have overcome this by introducing a novel \textit{``series" distance measure}. The distance between classes next to each other is determined as before, i.e. $d_{k,k+1}=|f_k-f_{k+1}|+\alpha |c_k-c_{k+1}|$. Note that $\alpha |c_k-c_{k+1}|$ can be simplified to the average frequency $\bar{f}$. The distance between two classes $s$ and $t$ with $s<t$ is determined by adding up previous distances:
\begin{equation} 
  d_{st} = \sum_{k=s}^{t-1} d_{k,k+1}. \label{eq:series distance}
\end{equation}
This can be expressed as a recursion $d_{k,k+2}=d_{k,k+1}+d_{k+1,k+2}$. All individual series distances form a corresponding matrix $D=(d_{st})$. Example \ref{fig:Example}'s histogram has similar characteristics to those observed in the case study, i.e. classes with low demand occur more frequently than those with high demand (a type 1 MTS-MTO hybrid).
These distance measures are used to build an agglomerative hierarchical cluster tree. We use the Unweighted Pair Group Method with Arithmetic Mean (UPGMA) to build a dendogram (binary tree) \citep{Sokal1958}. 
\begin{example}[Dendogram] \label{ex:dendogram}
We continue the previous example and build an agglomerative hierarchical cluster tree (see figure \ref{fig:Example} (b)). The average distance between class 1 and 2 is 62.5. The average distance to class three from one and two is $\frac{\breve{d}_{13}+\breve{d}_{23}}{2} = 73.8$. This cluster represents MTS items. Class 5 is in between the MTS and MTO cluster. 
\end{example}
In general the distance of two clusters $S$ and $O$ are determined by the average pair distances:
\begin{equation}
\frac{1}{n m}\sum_{s \in S}\sum_{o \in O} d_{s,o},
\end{equation}
where $n$ and $m$ are the number of elements in $S$ and $O$ respectively.
A possibility to improve the distance measures further is by considering holding costs, setup costs and shortage costs.

The procedure developed above is summarised in algorithm \ref{alg:cluster detection}.
\begin{algorithm}
    \caption{MTS/MTO cluster detection \label{alg:cluster detection}}
    \begin{algorithmic}[1]
    \Require equidistant time series of demands $y$, no major innovation in time series
    \Ensure MTO threshold $m$, MTS threshold $M$
    
		\State remove linear trend from time series $y$ and obtain $\dot{y}$
		\State transform $\dot{y}$ into histogram with classes $c_i$ and corresponding frequencies $f_i$
		\State create series distance matrix $D=(d_{st})$ using equation (\ref{eq:series distance})
		\State determine hierarchical cluster tree using unweighted average distances (see example \ref{ex:dendogram}).
		\State extract three clusters from the tree, identifying MTO and MTS thresholds
    \end{algorithmic}
\end{algorithm}
This procedure offers a systematic way of identifying initial MTS and MTO thresholds.
It may be the case that a product has no MTO products. However, the above procedure will always identify a potential MTO class. In such cases we suggest to fall back to classic service level strategies. For instance ensuring that predicted demand is fulfilled in 95\% of the time periods.
We require that there are no ``major" innovations in the time series. However, that issue and a solution are addressed in the following section.
The first step is the remove of linear trend. This is a first approximation which can be followed with higher order trend removals in accordance to a Taylor series development.
Next a histogram is derived. We would like to suggest as future work to use Kernel Density Functions. \cite{Botev2010} have proposed a method which has the advantage of being free of an underlying normal distribution assumption.
Step 3 and 4 were already explained in this section.
The last step identifies the MTO and MTS thresholds by using the agglomerative hierarchical clustering that identifies three clusters. The introduced distance ensures that the clusters are adjacent rather than overlapping.

\subsection{Time series\label{ssec:time series}}
In the previous section we have introduced a technique to identify MTO items; assuming a hybrid MTS-MTO time series. Removing MTO items from the hybrid time series will improve subsequent time series analysis for the remaining MTS items. This in turn causes savings in the holding costs. The production schedule can be determined via solving the Economic Lot Scheduling Problem. Alternatively general time series analysis and forecasting techniques can be used to determine the anticipated demand and derive a production schedule.

Let us consider three theoretical cases of time series shapes: (1) constant demand; (2) linear increasing demand; and (3) demand undergoes an innovation. The first case leads to histograms with a single class of high frequency, which is identified as MTO or MTS. This is in accordance to our empirical expectations, i.e. either orders are known in advance or excellent predictability. That means no further information is required to make a decision. The second case leads automatically to case one after trend removal. Not removing the trend would equipartition the histogram, suggesting wrongly that higher demand consists out of MTO items. 
The third case is a change in trend (innovation), which linear trend removal would not cope with. However, an additional preprocessing step to the previously introduced procedure \ref{alg:cluster detection} is able to transform case three into case one or two.
Finding the innovation is achieved by using a weighted average intra-cluster distance measure (weighted pair group method with averaging, WPGMA) instead of the average measure. 
First the series distance (equation \ref{eq:series distance}) is determined for the time series $y$. This is followed by the WPGMA clustering and identifying two clusters.

The boundary between the two clusters identifies the innovation.
This is the cut-off point, that means we can use the remaining observations for clustering and identifying MTO items using algorithm \ref{alg:cluster detection}. Hence, we will improve the MTS demand predictions for the shortened time-series. Certainly the question remains open, what is a ``major break"? Our method divides the time series into two parts, which allows several approaches to find an answer. Two approaches could be used to answer the question: (1) comparing the angle between the trend lines; (2) expectation and variance considerations of the two parts.

\subsection{Summary}
An decision support system was developed that classifies MTO and MTS items. To be more specific the proposed method transforms a time into a histogram. The histogram is further processed using a cluster detection technique that identified  MTO and MTS categories. The success of this clustering is ensured by the proposed distance measure. 

In practice there are several issues, which still need to be dealt with. The next section will look at those and add a few additionally required techniques, which are better understood through the application of actual data.

\section{Case study results}

The background of the case study company was given in the introduction and elaborated on the fact that there are 250 products to be manufactured. Initially we will focus on one particular product, and later discuss a set of ten products which were produced on the same production line. This will make use of the methodology introduced in the previous section.
This section will give us a good understanding of the data, and introduce one more required technique motivated by the case study's practical requirements. 
This technique is an approach for practitioners and suggests an approach to dissolve the MTS or MTO ``grey zone".  

\subsection{Product categories}
\begin{figure*}
	\centering
	 \begin{tabular}{cc}
		\includegraphics[width=8cm]{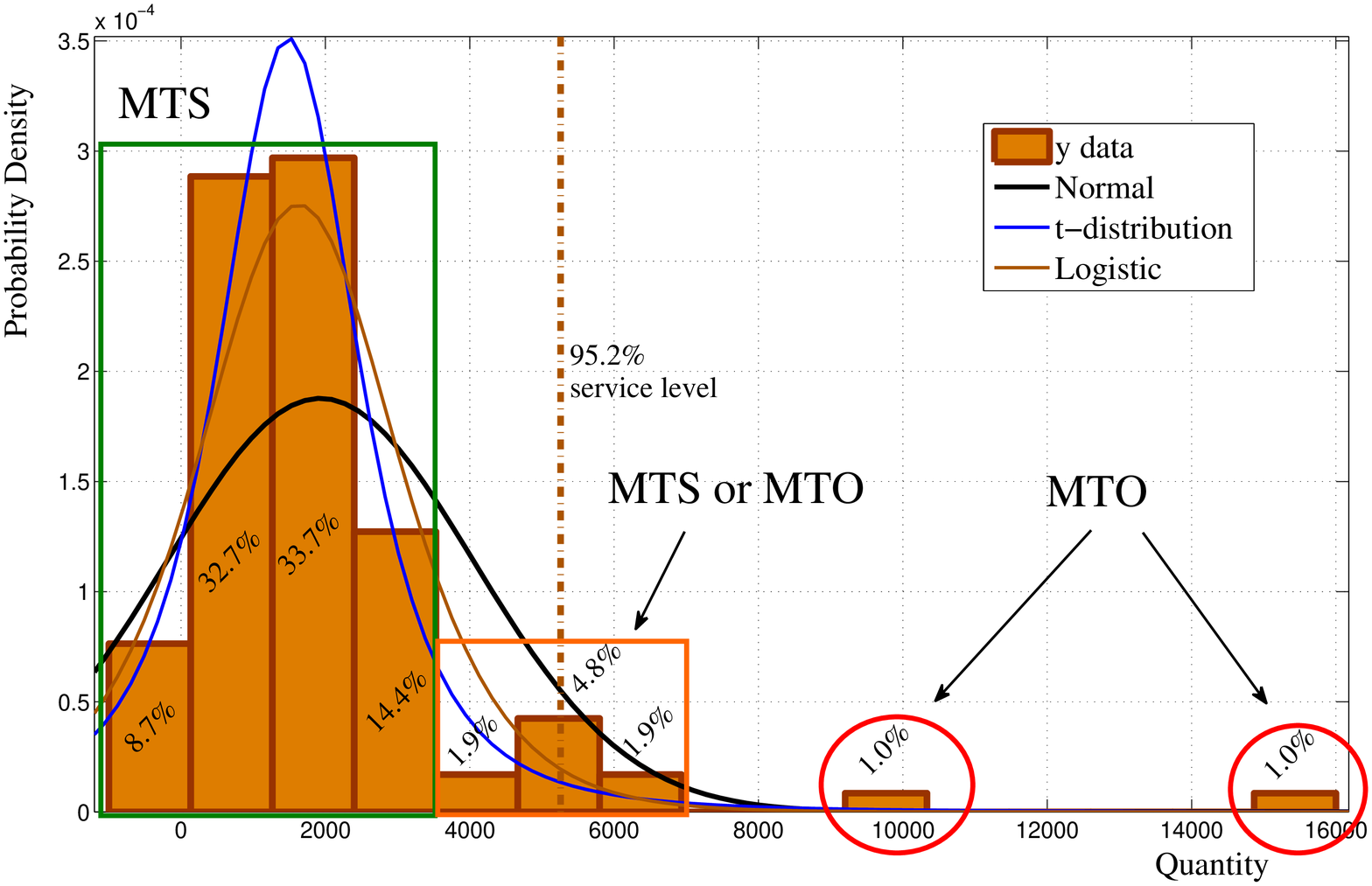} &
		\includegraphics[width=8cm]{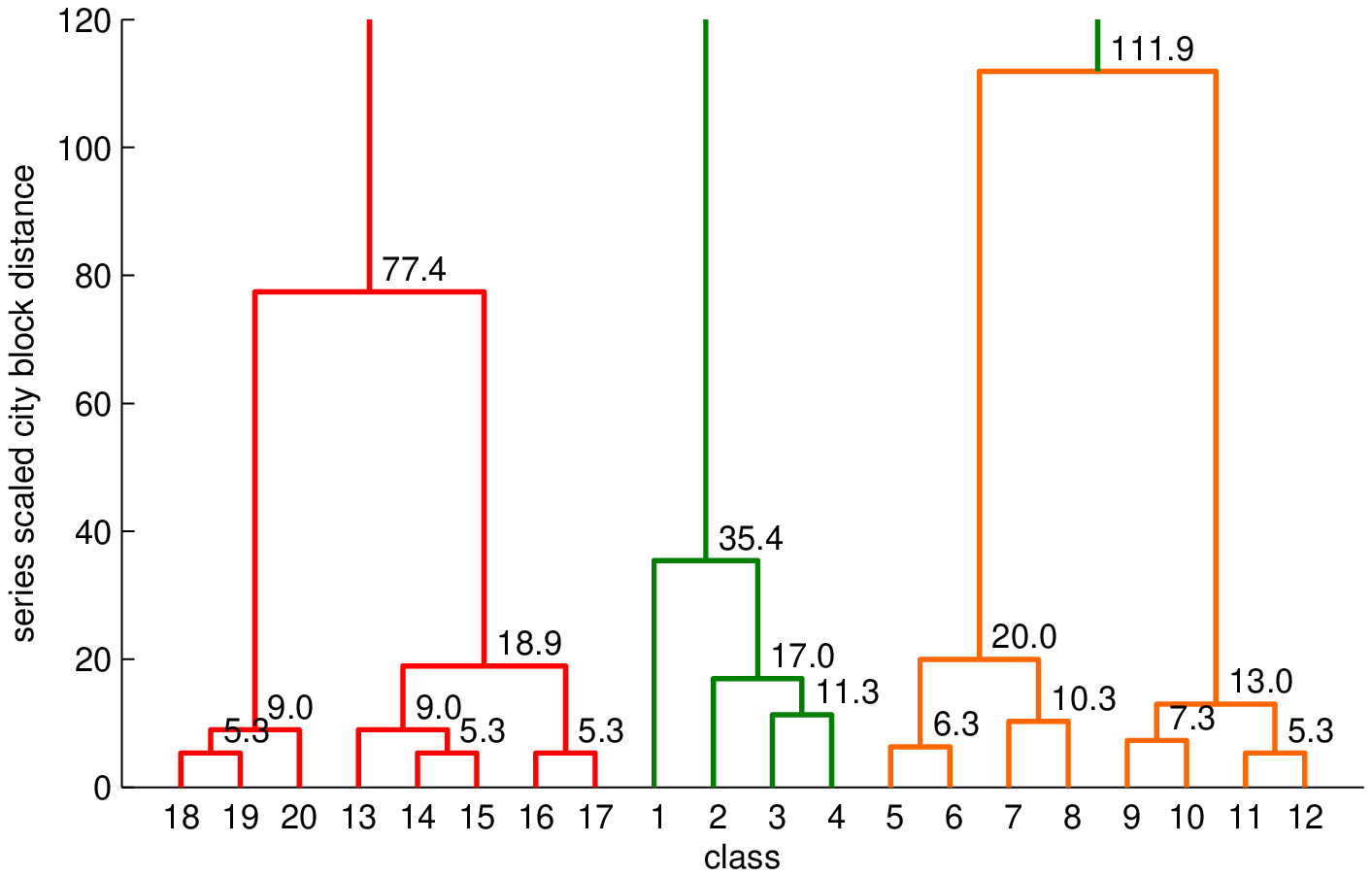} \\
		(a) & (b) 
	\end{tabular}
	 \caption{Product G01SL (a) Probability density and mass of demand identifying MTO and MTS, (b) Dendogram.}
	\label{fig:histo_MTS_MTO}
\end{figure*}

Figure \ref{fig:histo_MTS_MTO} (a) shows the demand profile for product G01SL over a period of two years ($n=104$ weeks).
It displays the frequencies of MTS/MTO items in 20 classes (bins).
This histogram was derived from the time series displayed in figure \ref{fig:TS-MTS} (a). 
The time series consists of equidistant demand observations $y=y_1,y_2,\dots,y_n$, i.e. $y_i$ is the number of items moved from the central warehouse (see section \ref{sec:company}) to the customer in week $i$. A high variability can be observed. A moving average of four weeks (black line) is shown and approximates a monthly demand profile. The linear trend (red line) shows that the demand is non-stationary.

The coefficient of variation $\frac{\bar{x}}{s}=1.1$ suggests a product being predominantly MTS. Here $\bar{x}$ is the sample mean and $s$ is the sample standard deviation. MTO items are visible by computing the $z$-score $z=\frac{x-\bar{x}}{s}$.
This identifies two weeks as outliers which have a $z$-score above 3.1. They are highlighted in figure \ref{fig:TS-MTS} (see dashed red lines). To be more precise week 73 and 82 have a $z$-scores of $z_{73}=6.5$ and $z_{82}=3.8$ respectively. That means for a normal distribution this would not be expected in 99.9\% of the cases. Despite the correct identification it should be noted that the Kolmogorov–Smirnov test does not confirm a normal distribution.

Considering figure \ref{fig:histo_MTS_MTO} the demand can be divided into three categories: MTO, MTS and a third one. Algorithm \ref{alg:cluster detection} identifies these categories using the hierarchical agglomerative clustering (see section \ref{sec:CDT}).
The clusters were created using the series city block distance (equation \ref{eq:series distance}). Figure \ref{fig:histo_MTS_MTO} (b) visualises and quantifies the intra cluster distances.
The MTO category occurs in two weeks with quantities 25,650 (12.8\%). MTS comprises 88.5\% weeks and a total quantity of 125,041 (62.3\%). The third category (MTO or MTS) happens in 9.6\% weeks with a quantity of 49,935 (24.9\%). 

Our objective is to get clarity about the third category. In order to achieve this we have to analyse the individual weeks.
We will use the symbol $W_s$ and $W_o$ to denote the MTS and MTO weeks respectively. The union of all weeks is abbreviated with $W$. That means the weeks within the MTS or MTO category are $W\setminus(W_w \cup W_o)$. The weeks can be derived using the MTO and MTS demand threshold determined by algorithm \ref{alg:cluster detection}.
Ultimately the intention is to end up with the MTS and MTO categories only. We can assume that the MTS category is pure with an average demand of 1,359 items per week. The two weeks identified as MTO occur in week 73 and 82 with demands of 15,700 and 9,950 respectively. 
A closer investigation of the MTO neighbours helps in deciding about which of the weeks $W_o^c$ should be added to the current MTO weeks $W_o$.
Week 73 has only one neighbouring week  with unusual high demand 5,096 (classified within the MTS or MTO category). The other neighbouring weeks $[69, 76]\setminus \{73,74\}$ have an average demand of 2,899.
A similar scenario is happening with week 82. Week 81 has a demand of 6,663 items. The surrounding 7 weeks window $[78, 85]\setminus \{81,82\}$ has an average demand of 1,542. The immediate neighbours of MTO from the MTS or MTO category have a demand volume of 5.9\%. 
We conclude that close to MTO identified weeks are likely to be part of MTO category. 
The pure MTS sample mean underestimates the MTO neighbouring demand and cannot be used for estimating the demand during the MTO weeks.
These heuristic considerations are used to decide about the final MTS and MTO classes.

\subsection{Final classification technique}
Based on the above considerations we will introduce a decision support mechanism to get even more clarity about MTS and MTO demand. We begin by using the third category (containing MTS or MTO weeks) as candidates for the MTO category. A candidate $a_i$ is the tuple $(w_i, y_i)$, where $w_i$ is the $i^{th}$ week and $y_i$ the corresponding demand. If a candidate is in the immediate neighbourhood (i.e.$w_{i-1}$ or $w_{i+1}$)  of a MTO identified week then the candidate becomes part of the MTO category. This is followed up recursively, i.e. a newly assigned MTO (ex-candidate) can cause candidates to be added to the MTO category.
The set of MTO-candidate weeks identified this way will be abbreviated with $W_o^c$.
For instance for product G01SL we only have two candidates, which are added to the MTO category. Now we are in the position to create our final demand profiles for MTO and MTS items.  
The final MTO weeks are $\hat{W}_o := W_o \cup W_o^c$ and the MTS weeks are $\hat{W}_s:=W \setminus \hat{W}_o$.
\begin{figure*}
	\centering
	 \begin{tabular}{cc}
		\includegraphics[width=8cm]{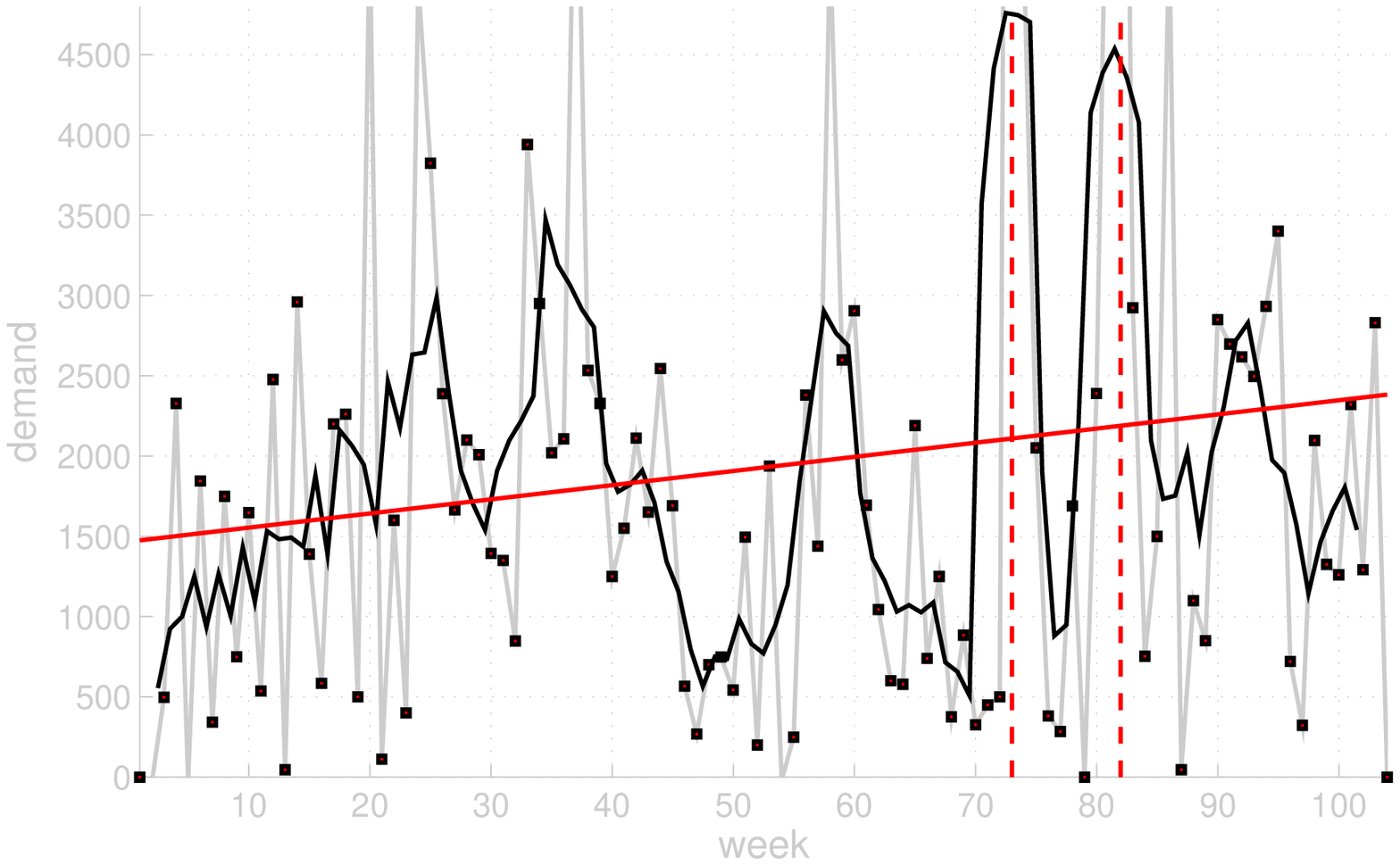} &
		\includegraphics[width=8cm]{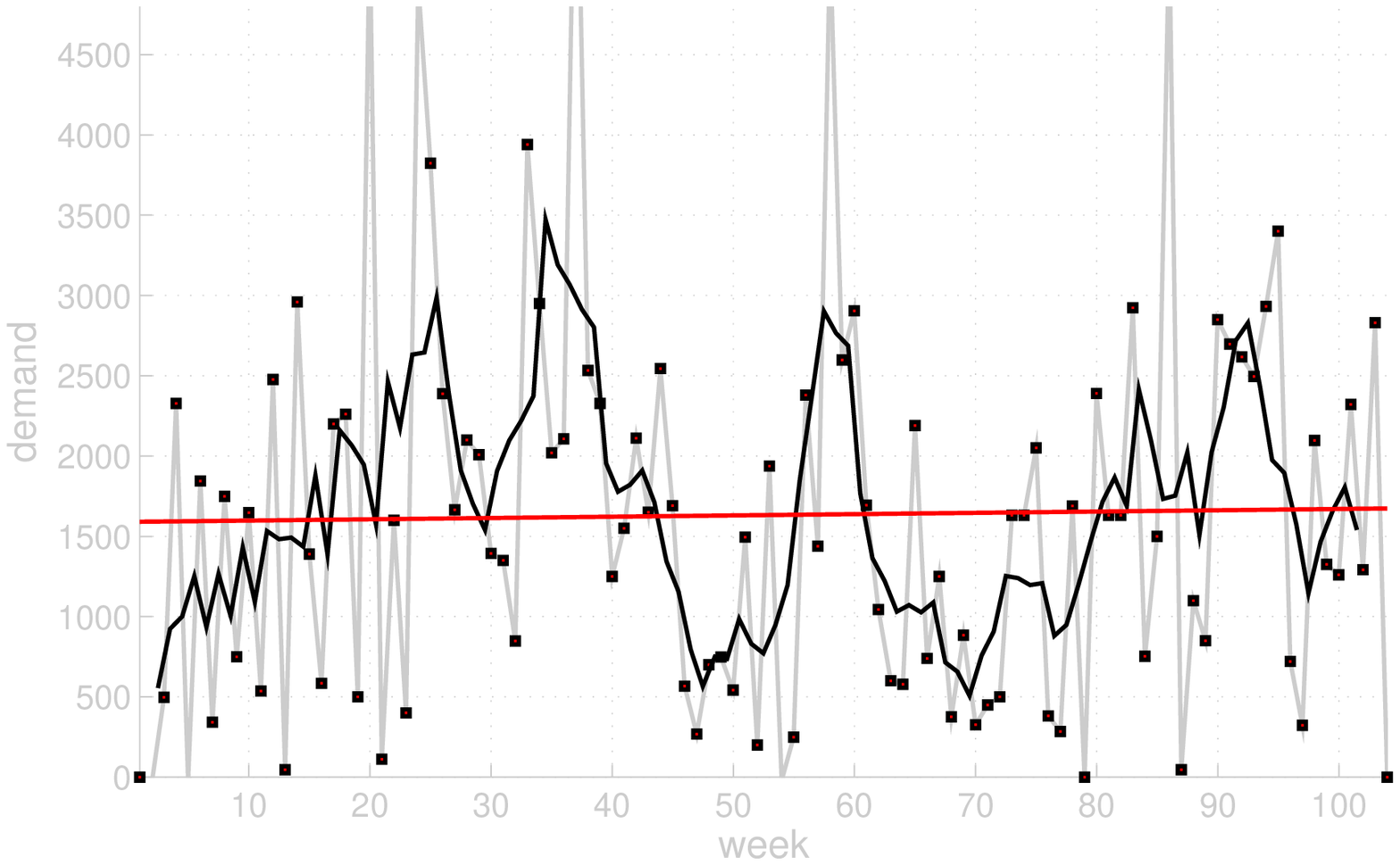} \\
		(a) & (b) 
	\end{tabular}
	 \caption{Product G01SL (a) hybrid time series, (b) MTS time series due to removal of MTO items.}
	\label{fig:TS-MTS}
\end{figure*}
We assume that weeks with MTO demand also have MTS demand, which is equal to the average MTS demand $\bar{D}_s$. 
Replacing the demand of the MTO weeks $\hat{W}_o$ with $\bar{D}_s$ gives us the time series displayed in figure \ref{fig:TS-MTS}, which we abbreviate with $y^s$.
\begin{equation}
 y_i^s 	= \begin{cases} y_i 				& \mbox{if } i\in W \setminus \hat{W}_o \\ 
												\bar{D}_s,  & \mbox{if } i\in \hat{W}_o \end{cases}
\end{equation}
The other time series is obtained by taking the demand during the MTO weeks $\hat{W}_s$ and reduce it by $\bar{D}_s$.
It is zero except for the MTO weeks and its recursively added candidates. 
\begin{equation}
 y_i^o 	= \begin{cases} 0 				& \mbox{if } i\in W \setminus \hat{W}_o \\ 
												y_i-\bar{D}_s,  & \mbox{if } i\in \hat{W}_o \end{cases}
\end{equation}
Of course $y_i^s+y_i^o$ is equal to $y_i$.

Considering product G01SL the final MTS demand is 84.6\%, which is an increase of 22.3\% to the pure MTS cluster demand. The MTO component has an apparent increase by 2.6\% resulting in a demand of 15.4\%. We use the word ``apparent" because the MTO cluster demand contains demand, which we have estimated to be $\bar{D}_s$ per week.

In summary this additional assignment technique has created a time series for MTS units and one for MTO units. 

\subsection{Practitioner approach and ten products}
\begin{figure*}
	\centering
		\includegraphics[width=1.00\textwidth]{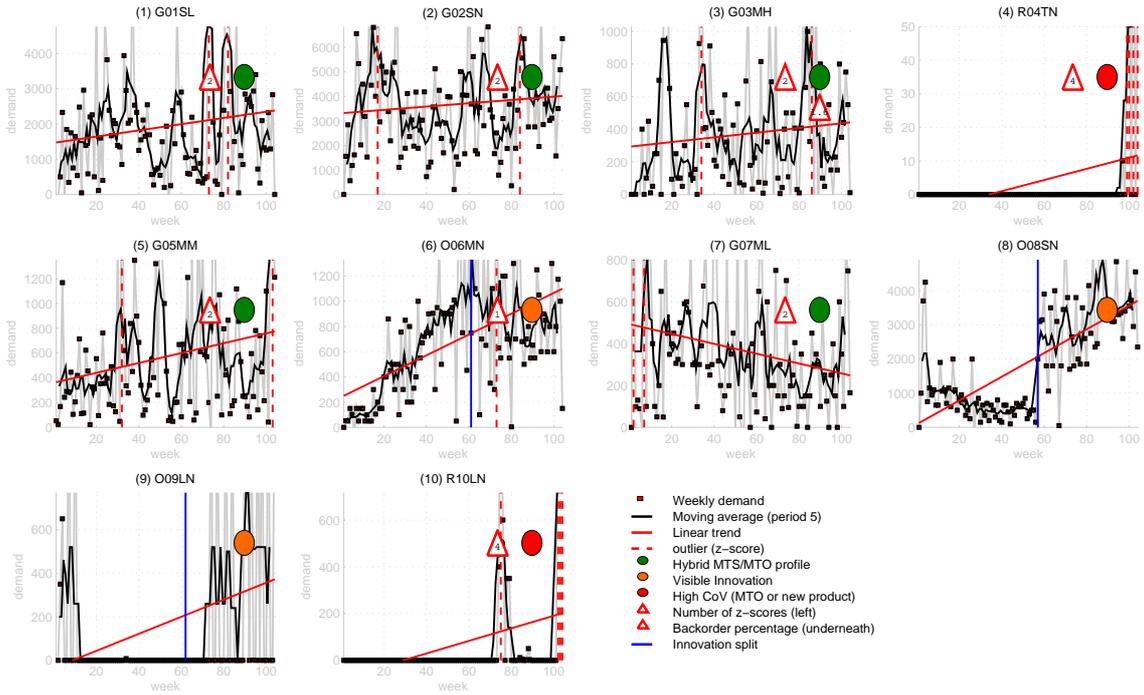}
	\caption{Time series showing the weekly demand.}
	\label{fig:TS-ten-products-comments}
\end{figure*}

We will give a guideline for practitioners to decide about MTO and MTS products/items in the absence of programming tool. These guidelines were applied to ten products. The timeseries for the ten products produced on the same manufacturing line are displayed in figure \ref{fig:TS-ten-products-comments}. Results in regards to savings and CoV are discussed in this context. Here savings are due to reduced holding cost, because the production knows the fraction of MTO units. The argument is that MTO items can be produced just in time. That means an improved planning process is possible.

The approach for practitioners is shown in algorithm \ref{alg:jump}. 
\begin{algorithm}
    \caption{Approach for practitioners \label{alg:jump}}
    \begin{algorithmic}[1]
    \Require Time series of demands $D$
    \Ensure MTO threshold $m$, MTS threshold $M$
    
		\State Analyse time series 
		\If {no break in structure (innovation) ``visible"}
			\State Identify MTO and MTS thresholds by
			\State Create histogram with frequencies $f_i$ and quantity centres $c_i$.
			\State Determine deltas $\Delta$ between consecutive frequencies $f_{1:n-1} - f_{2:n}$.
			\State MTS threshold $M$ is the quantity after the maximal frequency change $\max \Delta$ 
			\State MTO threshold $m$ is the quantity after the next maximal frequency change
		\ElsIf {break in structure identified}
			\State additional analysis required (set traffic light to orange)
		\ElsIf{coefficient of covariation high}
			\State MTO likely (alternatives: new product, product discontinuation) set traffic light to red
		\EndIf
    \end{algorithmic}
\end{algorithm}
As a first step the CoV should be used. Table \ref{tab:savings} gives the CoV for the ten products. Two of them reveal a CoV significantly above the average (product 4 and 10). The CoV was determined following the removal of the linear trend as discussed in section 4.1. Typically the removal of trend reduces the CoV, that means without removal a identification of MTO products is less reliable. Closer investigation of the two products reveals one as a new product being introduced and the other one as a pure MTO product with low demand. We will identify them with a red traffic light, in table 1, to indicate that no further investigation into these two products will take place (see algorithm \ref{alg:jump}, third \textbf{if} clause).

As a second step the practitioner can visually identify those time series showing an obvious innovation such as a break in structure or change in trend. These are marked with an orange traffic light in table \ref{tab:savings}. In our case study product six, eight and nine were identified. Algorithm \ref{alg:jump} only suggests that further analysis is required. In subsection \ref{ssec:time series} we introduced a time series clustering technique that allows to find the innovation for such time series. Using the new method allows the remainder of the time series to be analysed. Product six and eight remaining demand profiles were identified as hybrid MTS-MTO. Their estimated savings due to separate MTS and MTO production are 8.9\% and 15.0\%. Product nine is classified a MTO and no savings are possible.

\begin{figure*}
	\centering
		\includegraphics[width=1.00\textwidth]{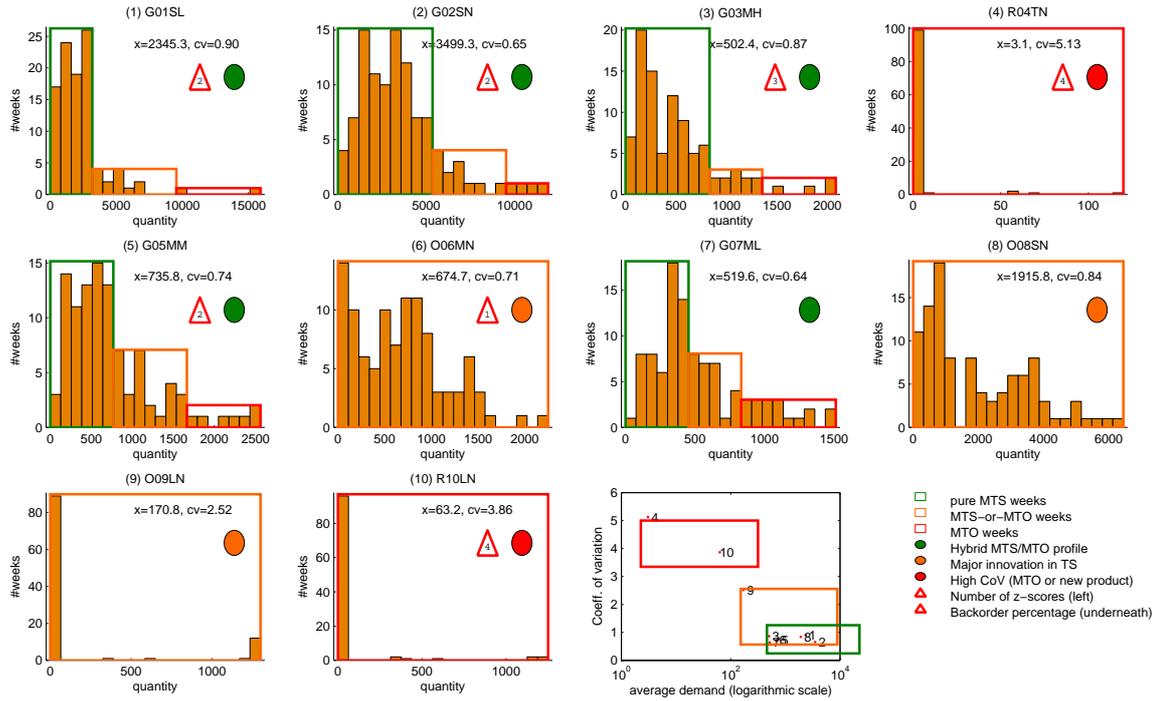}
	\caption{histograms demonstrating the MTS, MTS or MTO and MTO identification (trend removal when green traffic light).}
	\label{fig:Histos-ten-products}
\end{figure*}

The third step deals with the hybrid MTS-MTO character of the demand profiles. It should be noted that the proposed procedure omits or changes steps addressed in the original algorithm \ref{alg:cluster detection}. The major changes are the trend removal and the clustering technique. These adaptations lead to differences in the savings of about 5\%.
The results obtained with algorithm \ref{alg:cluster detection} are shown in table \ref{tab:savings} and \ref{tab:ten products MTS and MTO}. As we can see the savings are 18.1\% on average. This is due to the reduced holding costs by assuming neglect able holding costs for MTO items. That means MTO items are assumed to be produced just-in-time.
Figure \ref{fig:Histos-ten-products} visualises the MTS/MTO categories. The last sub-plot displays the coefficients of variation and their groupings.
The first table shows that the estimated savings are in the range of 13.4\% and 27.0\%. 
\input{Results1} 
Furthermore the CoV reduced in all cases (compare CoV and CoV-S), which will result in additional savings due to the reduced variation and increased production stability. Here, CoV-S is the abbreviation for the coefficient of variation for the MTS products, which varies between .77 and 1.07. CoV-O is the CoV for MTO products and varies between 2.96 and 5.70.
A technicality should be noted; there are weeks with negative demand (back-orders, BO). In particular product three had a negative demand of 1.65\%. For the clustering technique to work effectively back-orders above 0.3\% were treated separately. We have displayed the number of weeks as outlier via the $z$-score method in the last column of table \ref{tab:savings}. These numbers are equal or below the MTO weeks which were identified via algorithm \ref{alg:cluster detection}. The $z$-score assumes a normal distribution; however, all ten products failed the Kolmogorov-Smirnov and the Jarque-Bera test for normality. Many outlier detection methods depend on the normal distribution (see section \ref{sec:outlier techniques}). The empirical data cannot support their relevance for this case study. This is in accordance with time series literature in general, which attempts to find as a last remainder normally distributed noise.
Table \ref{tab:ten products MTS and MTO} shows the actual number of items per category. 
\input{Results2} 
Clearly indicating that product one and two have significantly more demand than the other products. The table shows first the absolute total demand over two years (except for the products with orange traffic lights). These numbers were obtained using algorithm \ref{alg:cluster detection}. They are also represented as volume percentage.
Using the recursive immediate neighbour approach the MTS or MTO category was dissolved resulting in the ``final" two MTS and MTO categories. An average of 81.9\% MTS items were found in the hybrid products. The last column of this table shows the average absolute number of MTS items in the final MTS category. 

\subsection{Comments}
The food processing sector continues to face increasing product volatility and service heterogeneity altering the historical demand profiles of SKU's. Scheduling and controlling products on the basis of a pure MTS approach has found in the case study firm to be futile and costly. Historically categorised MTS products regularly exhibited characteristics that align with a MTO approach rendering the production schedule redundant. Rush and exceptionally large orders were becoming the norm for the case study firm and were found to be difficult to manage through standard planning and control procedures.   
The methodology outlined in this section represents the modeling, analysis and subsequent implementation of a new way of planning and controlling products that demonstrate duality in their product categorisation. The algorithm offers practitioners the possibility to evaluate products and demand in terms of the production approach to be utilised to satisfy demand. The methodology above provides a support tool for decision-making and scheduling of production at the SKU level that separates orders in terms of MTO and MTS.The result for food processing firm showed that a separate production of the MTO items led to a reduction of holding costs by 18.1\% on average.

\section{Discussion and conclusion}

This empirical research has demonstrated how to deal with the combined MTS-MTO situation in the context of finite capacity. To date the authors are not aware of the existence of a published quantitative method which identifies MTO and MTS items automatically within a hybrid demand time series. The paper has reported on the challenges and difficulties of splitting MTO and MTS demand profiles to enable firms to manage fluctuations in customers demand and profile. The developed methodology provides practitioners with the tools to revise their order history and correlate it with the demand time series. Specifically the method supports the identification and scheduling of change in product status between MTS, MTO and MTO/MTS that could improve customer service and reduce costs. The new methodology has the potential to improve the performance of production planning and control within the food processing sector. The algorithm developed for practitioners is the first step in aiding production planners in achieving a stable schedule when faced with products exhibiting shifting categorisation status. 

The developed methodology is a valuable contribution to the MTO-MTS literature as it addresses, empirically as well as mathematically, a problem that planners face on a regular basis. The interaction effects of MTS-MTO demand characteristics with finite capacity is an area that had been identified by several researchers as an important area for research in planning and control. The data analysis from the study of ten products paints a complex picture of the dualistic nature of products in the food processing sector and suggests that further research is required to develop a stronger link between the developed methodology and practice. Identifying the MTO or MTS status of demand and converting the information to develop a robust manufacturing schedule, in the context of a dynamic MTO/MTS environment, is an area for future investigation.  

Through the development and implementation of the new methodology two further research directions have emerged. MTO and MTS demands that are similar in distribution are a challenge that have still to be researched and resolved in terms of the model. Identifying, segregating and managing the overlap in distributions requires further data collection over a wider range of products and longer time-series.  Integrating the single demand series for a hybrid product into the enterprise resource planning (ERP) system is also highlighted as an area for future research. Given the wide application of ERP systems in the food processing sector linking the algorithm to the main tool of production planners is a vital next step in developing a pathway for successful application and implementation of the new methodology.

\bibliographystyle{wg-elsarticle-harv}
\bibliography{MTO}

\end{document}

%% file: Results1.tex
     \begin{table}
       \centering
       \caption{Savings for products due to MTO item identification.}
						\resizebox{\columnwidth}{!}{%
     \begin{tabular}{rrrr|rrr|rr}
     Product &  TL   &  Savings & \#weeks &  CoV  &  CoV-S &  CoV-O &  BO   & z \\
     \hline
     G01SL &  \bgreen & 15.4\% & 104   & 1.10  & 0.79  & 5.70  & 0.04\% & 2 \\
     G02SN &  \bgreen & 13.4\% & 104   & 0.63  & 0.50  & 3.38  & 0.00\% & 2 \\
     G03MH &  \bgreen & 22.6\% & 104   & 1.20  & 1.07  & 3.83  & 1.65\% & 2 \\
     R04TN &  \bred &       & 104   & 5.13  &       & 5.13  & 0.00\% & 4 \\
     G05MM &  \bgreen & 27.0\% & 104   & 0.98  & 0.77  & 2.96  & 0.09\% & 2 \\
     O06MN &  \borange & 8.9\% & 44    & 0.52  & 0.45  & 3.66  & 0.00\% & 0 \\
     G07ML &  \bgreen & 24.2\% & 104   & 0.93  & 0.77  & 2.96  & 0.02\% & 2 \\
     O08SN &  \borange & 15.0\% & 48    & 0.40  & 0.32  & 1.98  & 0.00\% & 0 \\
     O09LN &  \borange &       & 43    & 1.73  & 4.86  & 1.73  & 0.00\% & 0 \\
     R10LN &  \bred &       & 104   & 3.86  &       & 3.86  & 0.00\% & 4 \\
     \hline
     \textit{} & \textit{Average} & \textit{18.1\%} & \textit{86.3} & \textit{1.65} & \textit{1.19} & \textit{3.52} & \textit{0.18\%} & \textit{1.8} \\
     \hline
     \end{tabular}%

				}
       \label{tab:savings}%
     \end{table}%

%% file: Results2.tex
     \begin{table*}
       \centering
       \caption{MTS and MTO volume per category.}
			\resizebox{\textwidth}{!}{%
     \begin{tabular}{rrrrr|rrr|rr|rr|r}
     Product &  TL   &  MTS  &  MT-S/O &  MTO  &  MTS\% & MT-S/O\% &  MTO\% &  fMTS &  fMTO &  fMTS\% &  fMTO\% &  mMTS \\
     \hline
     G01SL &  \bgreen &   125,041  &     49,935  &     25,650  & 62.3\% & 24.9\% & 12.8\% &     169,746  &     30,880  & 84.6\% & 15.4\% &     1,632  \\
     G02SN &  \bgreen &   183,955  &   145,144  &     52,282  & 48.2\% & 38.1\% & 13.7\% &     330,254  &     51,127  & 86.6\% & 13.4\% &     3,176  \\
     G03MH &  \bgreen &     12,088  &     17,737  &       8,416  & 31.6\% & 46.4\% & 22.0\% &       29,582  &       8,660  & 77.4\% & 22.6\% &        284  \\
     R04TN &  \bred &       &       &          320  &       &       & 100.0\% &       &          320  &       & 100.0\% &  \\
     G05MM &  \bgreen &     19,246  &     22,581  &     17,401  & 32.5\% & 38.1\% & 29.4\% &       43,224  &     16,004  & 73.0\% & 27.0\% &        416  \\
     O06MN &  \borange &     15,714  &     18,752  &       4,202  & 40.6\% & 48.5\% & 10.9\% &       35,217  &       3,452  & 91.1\% & 8.9\% &        800  \\
     G07ML &  \bgreen &     13,161  &     14,695  &     10,573  & 34.2\% & 38.2\% & 27.5\% &       29,138  &       9,291  & 75.8\% & 24.2\% &        280  \\
     O08SN &  \borange &     35,463  &     96,268  &     23,550  & 22.8\% & 62.0\% & 15.2\% &     132,044  &     23,237  & 85.0\% & 15.0\% &     2,751  \\
     O09LN &  \borange &       &       &     14,150  &       &       & 100.0\% &       &     14,150  &       & 100.0\% &  \\
     R10LN &  \bred &       &       &       6,576  &       &       & 100.0\% &       &       6,576  &       & 100.0\% & \\
     \hline
     \textit{} & \textit{Average} & \textit{   57,810 } & \textit{   52,159 } & \textit{   16,312 } & \textit{38.9\%} & \textit{42.3\%} & \textit{43.2\%} & \textit{   109,886 } & \textit{   16,370 } & \textit{81.9\%} & \textit{42.7\%} & \textit{   1,334 } \\
     \hline
     \end{tabular}}%
       \label{tab:ten products MTS and MTO}%
     \end{table*}%